\documentclass[11pt]{article}
\usepackage{amsmath,amssymb,color,graphics,epsfig}

\usepackage{amsfonts}

\newcommand{\hoch}[1]{$\, ^{#1}$}

\newcommand{\be}{\begin{equation}}
\newcommand{\ee}{\end{equation}}
\newcommand{\bea}{\setlength\arraycolsep{2pt} \begin{eqnarray}}
\newcommand{\eea}{\end{eqnarray}}

\def\ft#1#2{{\textstyle{\frac{\scriptstyle #1}{\scriptstyle #2} } }}
\def\fft#1#2{{\frac{#1}{#2}}}

\def\0{{\sst{(0)}}}
\def\1{{\sst{(1)}}}
\def\2{{\sst{(2)}}}
\def\3{{\sst{(3)}}}
\def\4{{\sst{(4)}}}
\def\5{{\sst{(5)}}}
\def\6{{\sst{(6)}}}
\def\7{{\sst{(7)}}}
\def\8{{\sst{(8)}}}
\def\sst#1{{\scriptscriptstyle #1}}

\begin{document}

\begin{flushright}
\hfill{KIAS-P12068}
\end{flushright}

\vspace{25pt}
\begin{center}
{\large {\bf Exact Green's Function and Fermi Surfaces from Conformal
Gravity}}

H. L\"u\hoch{1} and Zhao-Long Wang\hoch{2}

\vspace{10pt}

\hoch{1}{\it Department of Physics, Beijing Normal University,
Beijing 100875, China}

\vspace{10pt}

\hoch{2} {\it School of Physics, Korea Institute for Advanced Study,
Seoul 130-722, Korea}

\vspace{40pt}

\underline{ABSTRACT}
\end{center}

We study the Dirac equation of a charged massless spinor on the general charged AdS black hole of conformal gravity. The equation can be solved exactly in terms of Heun's functions. We obtain the exact Green's function in the phase space $(\omega,k)$. This allows us to obtain Fermi surfaces for both Fermi and non-Fermi liquids.  Our analytic results provide a more elegant approach of studying some strongly interacting fermionic systems not only at zero temperature, but also at any finite temperature.  At zero temperature, we analyse the motion of the poles in the complex $\omega$ plane and obtain the leading order terms of the dispersion relation, expressed as the Laurent expansion of $\omega$ in terms of $k$. We illustrate new distinguishing features arising at the finite temperature.  The Green's function with vanishing $\omega$ at finite temperature has a fascinating rich structure of spiked maxima in the plane of $k$ and the fermion charge $q$.

\vfill {\footnotesize Emails: mrhonglu@gmail.com;\ \ \
zlwang@kias.re.kr}

\thispagestyle{empty}





\newpage

\noindent{\bf \large Introduction}: Charged black holes that are asymptotic to anti-de Sitter spacetimes (AdS) play important roles in applying the AdS/CFT correspondence to study some traditional condensed matter physics, including superconductors and Dirac-Fermi systems.  This paper reports that the analytic Green's functions in the momentum space of some strongly interacting Dirac-Fermi systems can be obtained from charged AdS black holes of conformal gravity.

It was proposed that such a black hole is dual to some strongly coupled fermionic system at finite charge density such as non-Fermi liquids \cite{Lee:2008xf}.  The procedure of deriving the Green's function was spelled out in \cite{lmv,flmv}, where the Dirac equation of charged spinor in the extremal Reissner-Nordstr\"om black hole was studied. Imposing the appropriate horizon boundary condition on the solution, one can read off the Green's function in the momentum space $(\omega, k)$ from its asymptotic behaviour. Unfortunately, there is no analytic solution for the Dirac equation, and the numerical approach was adopted in \cite{lmv,flmv}.  Nevertheless, the Fermi surfaces, which are defined as poles of the Green's function with vanishing $\omega$, were determined, and the Green's function for small $\omega$ on the Fermi surface was shown to take the form \cite{lmv,flmv}:
\begin{equation}
G(\omega,k)=-\fft{h_1}{(k-k_F)-v_F^{-1}\omega-h_2e^{{\rm i}\gamma_{k_F}}
\omega^{2\nu_{k_F}}}\,,\label{genfssmallw}
\end{equation}
where $v_F$ is the Fermi velocity, $h_1, h_2, \gamma_{k_F}, \nu_{k_F}$ are constants; they were determined numerically in \cite{lmv,flmv}.  The system describes non-Fermi liquids for $\nu_{k_F}<\fft12$. It turns out that indeed the case with $\nu_{k_F}<\fft12$ can arise from the charged AdS black hole as well as $\nu_{k_F}>\fft12$.

While physics can be adequately studied numerically, it is more elegant and satisfying if we can find an analytical result.  Recently, two equally-charged AdS black hole in $D=5$ gauged supergravity was considered in \cite{gure}.  In some suitable extremal limit, the Green's function with $\omega=0$ can be solved exactly and the Fermi surfaces $k_F$ can be determined analytically.  A procedure \cite{lmv,flmv} can then be used to derive the Green's function for small $\omega$ and the analytical expressions of the constants in (\ref{genfssmallw}) were obtained \cite{gure}.  However, the black hole of this example suffers from having a curvature singularity that coincides with the horizon.  There also lacks an analytic expression for general $G(\omega,k)$ and furthermore there is no analytic result for non-extremal black holes.  The black holes considered in \cite{lmv,flmv,gure} are arguably among the simplest ones in usual two-derivative gravities or supergravities. It is thus unlikely to find new black holes without convoluted matter in these theories whose Dirac equation becomes exactly solvable.

\noindent{\bf \large Charged black hole in conformal gravity}: Charged AdS black holes can also arise in conformal gravity. Conformal pure gravity in four dimensions is a four-derivative theory constructed from the Weyl-squared term.  Its minimum coupling to the Maxwell field preserves the conformal symmetry.  The Lagrangian is given by
\begin{equation}
e^{-1}{\cal L} =  \ft12\alpha C^{\mu\nu\rho\sigma}
C_{\mu\nu\rho\sigma} + \ft13\alpha F^2\,,\label{conflag}
\end{equation}
where $F=dA$. The Weyl-squared term is crucial in constructing critical gravity \cite{lpcritical} and supergravity \cite{lpsw}. It was argued that Einstein gravity can emerge from conformal pure gravity in the IR region \cite{maldconf}.  Furthermore, conformal gravity can be supersymmetrised in the off-shell formalism \cite{ledu}.

   The most general spherically-symmetric black hole up to a conformal
transformation was obtained in \cite{Riegert:1984zz}.  For our purpose of studying the Dirac-Fermi system on a plane, we construct the torus-symmetric solution, given by
\begin{eqnarray}
ds^2 &=&-f dt^2 + \fft{dr^2}{f} + r^2 (dx^2 + dy^2)\,,\qquad
A=Q\Big(\fft{1}{r_0} - \fft{1}{r}\Big)dt\,,\cr 
f&=&\fft{(-\Lambda/3)}{r} (r-r_1)(r-r_2) (r-r_0)\,,\cr
Q^2&=&(r_1 r_2)^2 + (r_0r_1)^2 + (r_0r_2)^2 - r_1 r_2 r_0(r_1 + r_2 + r_0)\,.\label{rieg}
\end{eqnarray}
Without loss of generality, we let $r=r_0>0$ be the horizon. It follows that if $(r_1,r_2)$ are real, neither is bigger than $r_0$.  They can also form a pair of complex conjugates.   The thermodynamics of the black hole is analysed in \cite{lllw}.  Note that the solution contains three non-trivial parameters, associated with the mass, charge and the massive spin-2 hair.  We set $\Lambda=-3$ so that the asymptotic AdS has unit radius.  The solution becomes extremal if we let $r_2=r_0$, (or $r_1=r_0$, but not both.) The solution of $r_1=0$ was obtained in \cite{luwangsusylif} by employing the pseudo-supersymmetry related to off-shell conformal supergravity.

\noindent{\bf \large Charged fermion and Green's function:}  We are now in the position of studying the Dirac equation for a charged massless spinor:
\begin{equation}
\gamma^\mu (\partial_\mu + \ft14\omega_{\mu}^{ab}\Gamma_{ab} - {\rm
i} q A_\mu)\Psi=0\,.\label{diracE0}
\end{equation}
We follow the procedure outlined in \cite{lmv,flmv}.  For static ansatz, the contribution of the spin connection in the Dirac equation
(\ref{diracE0}) can be absorbed by the scaling of the field
$\widetilde\Psi= (-g g^{rr})^{\fft14}\Psi$.  In the Fourier mode
$\widetilde \Psi \sim e^{-{\rm i} \omega t + {\rm i} k x}\hat \Psi$
with the momentum lying in the $x$ direction only, the Dirac equation reduces,
with suitable choice of $\Gamma$-matrices, to two decoupled equations for $\hat \Psi =(\hat \psi_1,\hat\psi_2)$.   The equation for $\hat \psi_2$ is the same as that for $\hat \psi_1$ up to $k\leftrightarrow -k$.  The $\hat \psi_\alpha$'s are themselves two-component spinors.  Let $\hat\psi_1=(u_1,u_2)^T$ and
$u_{\pm}=u_1\pm {\rm i} u_2$, we have
\begin{eqnarray}
&&u_+'+\bar{\lambda}_1(r)u_+=\bar{\lambda}_2(r)u_-\,,\qquad
u_-'+\lambda_1(r)u_-=\lambda_2(r)u_+\,,\label{twofo}\\
&&\lambda_1(r)=\fft{{\rm i} (\omega+q a)}{f}\,,\qquad
\lambda_2(r)=-\fft{{\rm i} k}{r \sqrt{f}}\,.
\end{eqnarray}
(Note that if we had considered the massive Dirac equation, we would
have instead $\lambda_2(r)=\frac{m}{\sqrt{f}}-\fft{{\rm i} k}{r \sqrt{f}}$.) It
follows that
\begin{eqnarray}
&&u_+''+\bar{p}_1(r)u_+'+\bar{p}_2(r)u_+ =0\,,\qquad
u_-''+p_1(r)u_-'+p_2(r)u_- =0\,,\\
&&p_1(r)=-\frac{\lambda_2'}{\lambda_2},\qquad
p_2(r)=|\lambda_1|^2-|\lambda_2|^2+p_1(r) \lambda_1+\lambda_1'.
\end{eqnarray}
There are two independent solutions for $\Psi$ and they can be organised as in-falling and outgoing modes on the horizon.  The nature of the back hole requires keeping only the in-falling mode, which is associated with the {\it retarded} Green's function.  To determine the Fermi surfaces, one only needs $G(\omega=0,k)$.  The horizon condition for the $\omega=0$ solutions is more subtle.  In extremal black holes, it turns out that one mode diverges on the horizon and it should be dropped whilst the other mode converges.  However, for non-extremal back holes, both modes are convergent. There is no rule to impose the horizon condition if one has only solutions with vanishing $\omega$.

Once the horizon condition is appropriately imposed, the Green's function can be read off from the asymptotic behaviour of the wave solution, given by
\begin{equation}
G_1 = - G_2^{-1} = \lim_{r\rightarrow \infty} \fft{u_2}{u_1} =
-{\rm i}\lim_{r\rightarrow\infty}\frac{u_+-u_-}{u_++u_-}\,.\label{gfformula}
\end{equation}
Here, $G_1$ and $G_2$ are diagonal entries of the 2 by 2 matrix of the Green's function associated with the massless spinor. In particular $G_1$ corresponds to the right-handed spinorial operator and the $G_2$ corresponds to the left, which we shall not consider in this paper.  In this ``standard'' quantization, the function $u_1$ is treated as a source and $u_2$ is the response. It is clear that $u_1$ and $u_2$ are symmetric in the wave function.  This leads to an alternative quantization in which the roles of $u_1$ and $u_2$ are reversed.  In this case, the Green's function of the right-handed spinor is given by $-G_2$ instead.

We find that the wave equations can be solved exactly for our general black hole (\ref{rieg}).  The solution can be expressed in terms of Heun's functions, which are too complicated for detailed discussions on the fine properties of the Green's function.  In what follows, we shall give some detail analysis for the simplest case: $r_1=0$ and $r_2=r_0$.  We shall then present the salient results in the increasing order of the complexity of the black hole and more detailed analysis are presented in \cite{lllw}.

\noindent{\bf \large Case 1: $r_1=0$ and $r_2=r_0$.} We begin with the simplest extremal solution, which was obtained in \cite{luwangsusylif}.
The general solutions for $u_\pm$ are
\begin{eqnarray}
u_-=\ft{c_1}{\sqrt{z}} W_{\kappa,\nu}(z) + \ft{c_2}{\sqrt{z}} W_{-\kappa,\nu}(-z)\,,\qquad
u_+=\ft{{\rm i} k c_1}{r_0\sqrt{z}} W_{-\kappa^*,\nu}(z) +
\ft{{\rm i} r_0 c_2}{k\sqrt{z}} W_{\kappa^*,\nu}(-z)\,,\label{upmsol}
\end{eqnarray}
where $c_1$ and $c_2$ are two integration constants, and $W_{\kappa,\nu}(z)$ denotes the Whittaker function. Further notations are specified as follows
\begin{equation}
\kappa = \ft12 + {\rm i} q\,,\qquad \nu = \sqrt{\ft{k^2}{r_0^2} - q^2}\,,\qquad
z=-\fft{2{\rm i} \omega\, r}{r_0(r-r_0)}\,,
\end{equation}
and $\kappa^*$ is the complex conjugate of $\kappa$.  As $r$ approaches the horizon $r_0$, we have $z\rightarrow -\infty$, and the Whittaker function behaves as $
W_{\kappa,\nu} \sim e^{-\fft12 z} z^{\kappa} (1 + {\cal O}(z^{-1}))\,.
$ Thus, the in-falling solution, selected by the black hole, is given by (\ref{upmsol}) with $c_2=0$.  The retarded Green's function can now be read off straightforwardly, given by
\begin{equation}
G_1=-G_2^{-1} = {\rm i} \fft{1-\gamma}{
1  +\gamma}\,,\qquad
\gamma = \fft{{\rm i} k\,U (1-{\rm i} q + \nu, 1 + 2\nu, -\fft{2{\rm i}\omega}{r_0})}{r_0\, U(-{\rm i} q + \nu, 1 + 2\nu, -\fft{2{\rm i}\omega}{r_0})}\,, \label{extrgamma}
\end{equation}
where $U$ is the confluent hypergeometric function, defined by
\begin{equation}
U(a,b,z) = \fft{\Gamma(1-b)}{\Gamma(a-b+1)} {}_1F_1 (a;b;z) +
\fft{\Gamma(b-1)}{\Gamma(a)} z^{1-b} {}_1F_1 (a-b+1;2-b;z)\,.
\end{equation}
We can now investigate the Fermi surfaces. For $\omega =0$, we find
\begin{equation}
G_1=-G_2^{-1} = -\sqrt{\fft{k+q r_0}{k-q r_0}}\,.
\end{equation}
Thus it becomes clear that $k_F= q r_0$ defines the only Fermi surface for the standard quantization and $k_F=-q r_0$ for the alternative quantization.  For both cases, we have $\nu_{k_F}=0$.  Such an extreme situation of non-Fermi liquids were also observed in extremal Reissner-Nordstr\"om black holes \cite{lmv,flmv}. Our general $G(\omega, k)$ allows us to study the behaviour in greater detail. Let us examine the behaviour of the Green's function near the Fermi surface, which occurs at $k_F=q r_0$. If $\omega/r_0$ approaches zero much faster than $(k-k_F)$, we find that
\begin{equation}
G_1 \sim  \fft{\sqrt{2 k_F (k-k_F)}}{(k-k_F) + q \sqrt{\fft{2(k-k_F)}{k_F}}\, \omega +
2(k-k_F)(-\fft{2{\rm i}\, \omega}{r_0})^{\sqrt{\fft{2(k-k_F)}{k_F}}}}\,.
\end{equation}
Comparing to the small-$\omega$ formula (\ref{genfssmallw}), we find
\begin{eqnarray}
v_F^{-1}\sim 0-q\sqrt{\ft{2(k-k_F)}{k_F}}\,,\qquad
h_1\sim 0-\sqrt{2k_F(k-k_F)}\,,\qquad h_2\sim 0-2(k-k_F)\,.
\end{eqnarray}
Analogous result can also be derived in the alternative quantisation \cite{lllw}.      It is instructive also to consider how the Green's functions behave for small $\omega$ after we have literally fixed $k=k_F$.  We find
\begin{equation}
G(\omega, k_F)\sim -{\rm i} + 2 q \Big(\log(-\fft{2{\rm i} \omega}{r_0}) -2\gamma -\chi({\rm i} q)\Big)\,,
\end{equation}
where $\gamma$ is the Euler number and $\chi$ is the digamma function.
The divergence is logarithmic in small $\omega$.

\begin{figure}[ht]\center
\includegraphics[width=7cm]{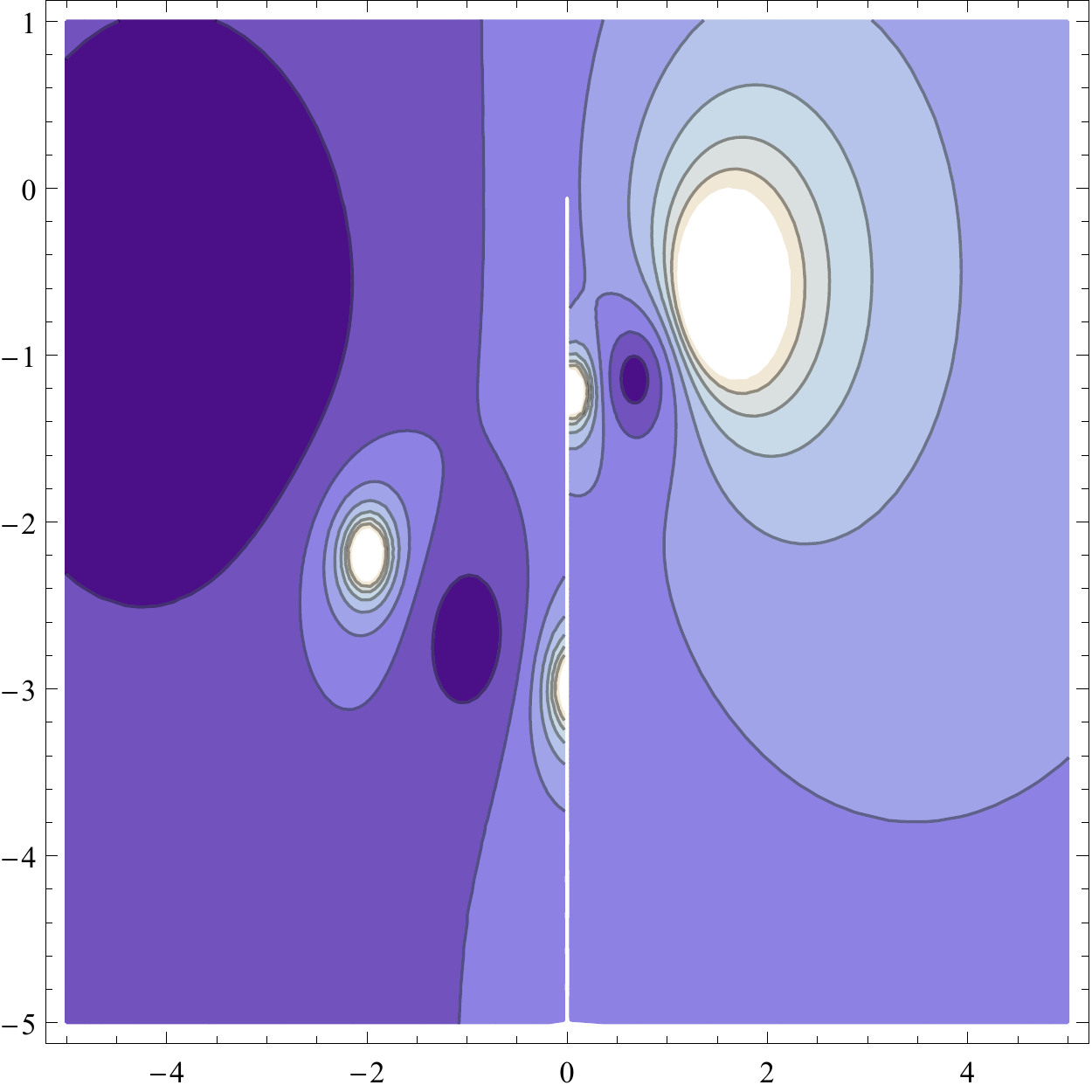}\ \ \ \
\includegraphics[width=7cm]{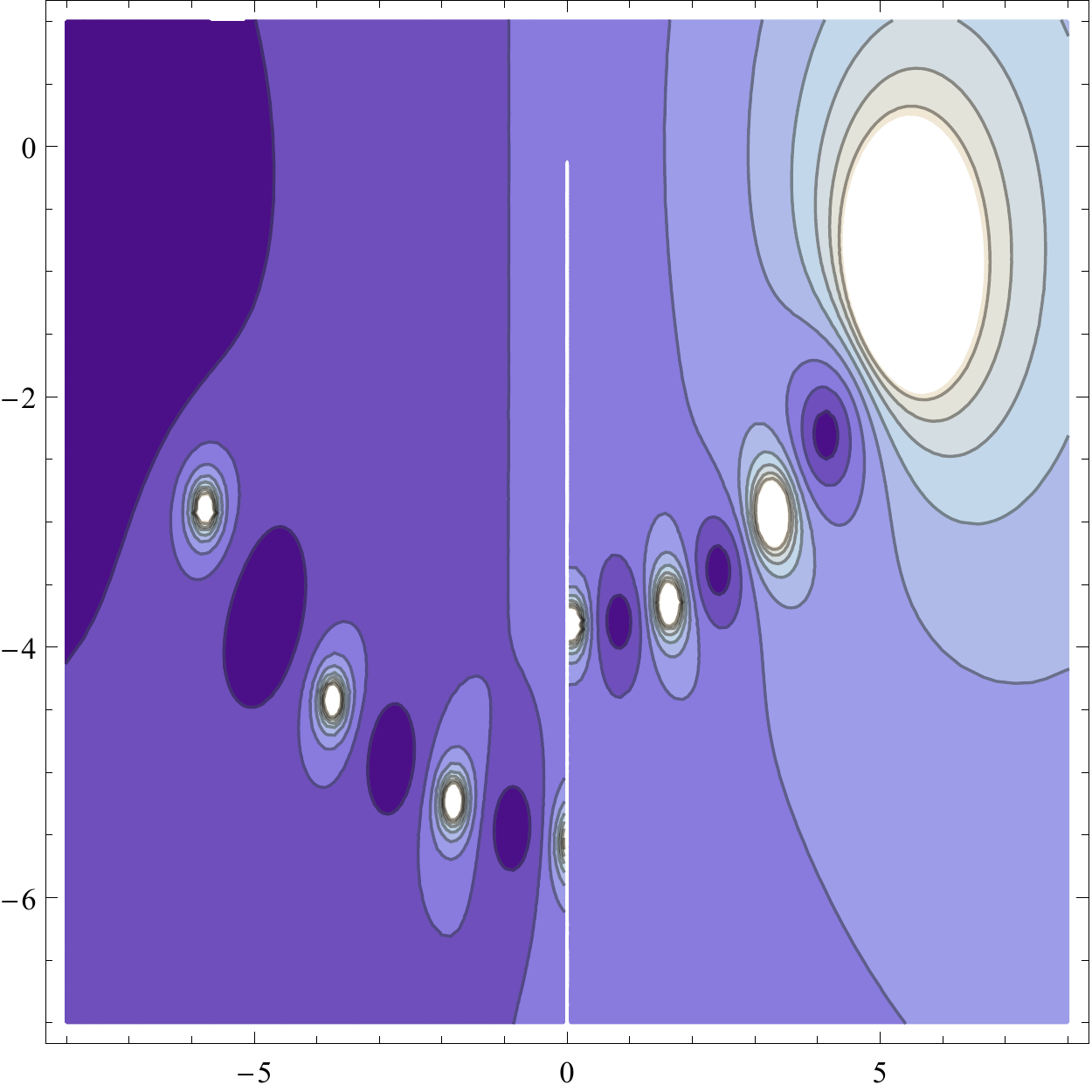}
\caption{The contour plots of $|G|$ in the plane of complex $\omega$.  All the poles (white dots) and zeros (dark dots) occur in the lower half plane.  Their number increases for larger $k$. A branch cut in the negative imaginary axis creates a shift to the curves.}
\label{contour1}
\end{figure}

    Our Green's function (\ref{extrgamma}) for general $\omega$ and $k$
enables us to find poles and even the dispersion relations in the large parameter regions of $\omega$ and $k$.  Let us consider $r_0=1$ and $q=1$.  In Fig.~\ref{contour1}, we give two contour plots of $|G|$ in the plane of complex frequency.  The left has $k=3$ and the right has $k=7$.  We see that the number of poles and zeros, corresponding to the white and dark dots, increases with larger $k$. There is a branch cut along the negative imaginary axis, and the right and left ``curves'' disjoin at the branch cut. Note that the gap between the left and right curves disappears when $q$ approaches zero.  It is also of interest to note that in the oscillating region with imaginary $\nu$, we find that the small area of the lower left plane with $|\omega|<<1$ contains an infinite number of zeros and poles.

\begin{figure}[ht]\center
\includegraphics[width=8cm]{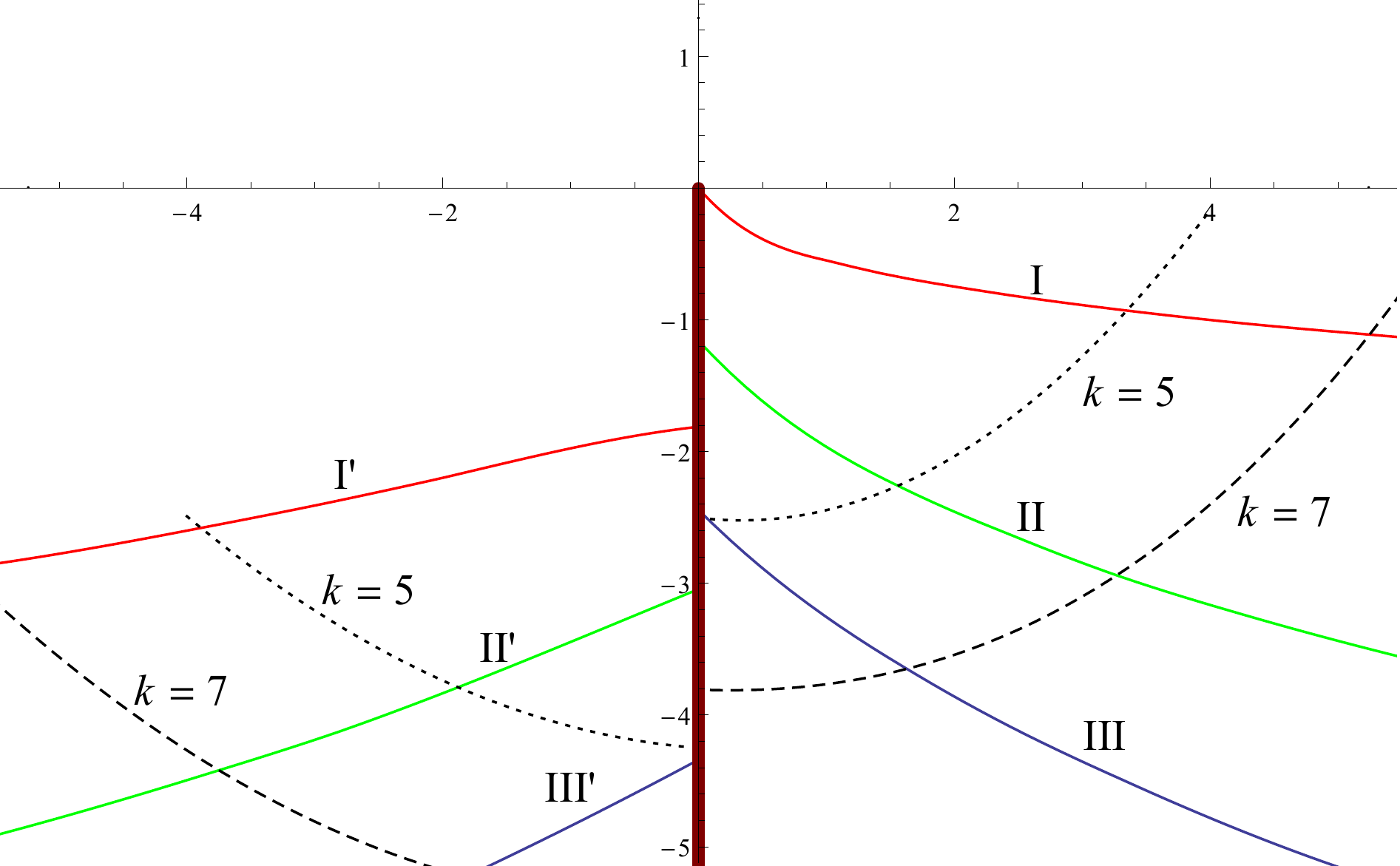}
\caption{This plot depicts the motion of poles in the plane of complex $\omega$. The solid lines are the locations of poles for continuous parameter $k$.  The dotted and dashed lines describe the artificial curves connecting the discrete poles for a given $k$, as in Fig.~\ref{contour1}.  The number of poles increases for larger $k$.  The negative imaginary axis is the branch cut. Analogous figure for the extremal Reissner-Nordstr\"om black hole, but only with $\omega/r_0<<1$,  was obtain in \cite{flmv}.}
\label{motion}
\end{figure}

We study a number of such contour plots and obtain the poles for a variety of $k$.  This allows us to draw the motion of poles in the complex $\omega$ plane in Fig.~\ref{motion}.  It is significantly different from the small-$(\omega/r_0)$ analysis \cite{flmv}, in which the artificial curve connecting the poles for constant $k$ is a straight line.

\begin{figure}[ht]\center
\includegraphics[width=5cm]{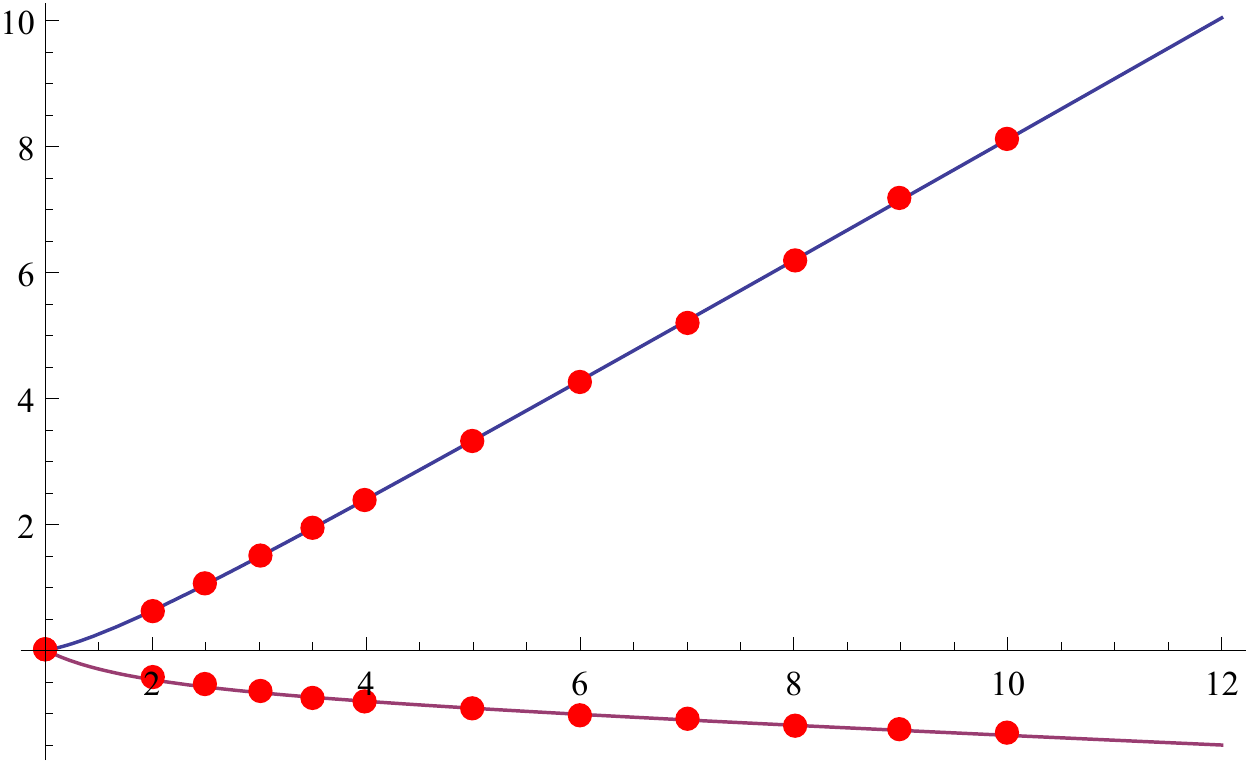}\ \ \ \
\includegraphics[width=5cm]{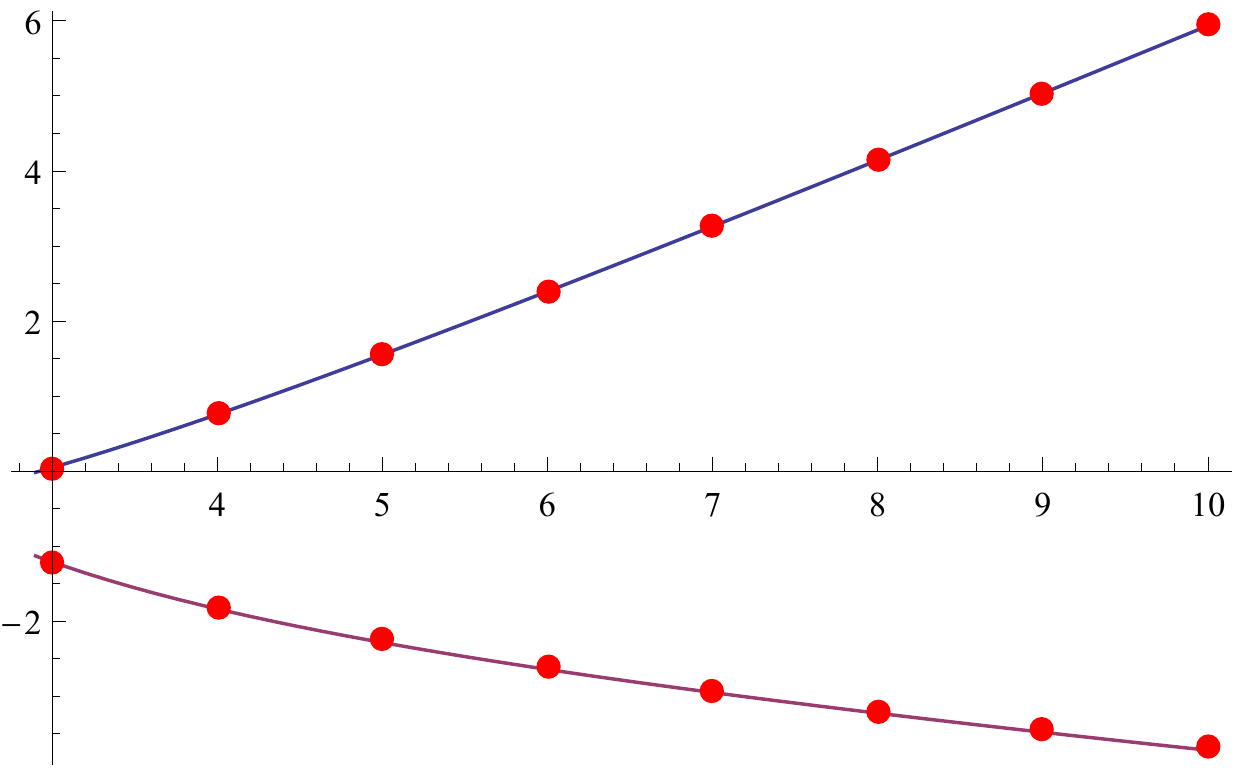}
\caption{The fitting of the dispersion relations (\ref{disperse}) (solid lines) and the actual poles (dots).  The left figure is for curve I and the right is for curve II.  In both cases, the horizontal axis is $k$; the upper line is ${\rm Re}(\omega)$ and the lower is ${\rm Im}(\omega)$.}
\label{disperseplot}
\end{figure}

The data associated with the solid lines in Fig.~\ref{motion} enable us to construct the dispersion relations for these curves.  We find that a very good fit is to use the Laurent series expansion of $\omega$  in terms of $k$:
\begin{equation}
\omega = \cdots + \fft{a_{-1}}{k} + a_0 + a_1 k + \cdots\,,\label{disperse}
\end{equation}
where $a_i$ are complex numbers.  For example, for the curves I, I$'$ and II, we have
\begin{eqnarray}
{\rm I}:&& a_{-1} = 0.68 + 0.81{\rm i}\,,\quad a_{0}=-1.65-0.73{\rm i}\,,\quad a_1=0.97 - 0.07{\rm i}\,,\cr
{\rm I'}:&& a_{-1}=0.05 + 0.05{\rm i}\,,\quad a_0=0.83-1.69{\rm i}\,,\quad a_1=-0.95 -0.18{\rm i}\,,\cr
{\rm II}:&& a_{-1}=2.65 + 5.36{\rm i}\,,\quad a_0 = -3.63 - 2.64{\rm i}\,,\quad
a_1=0.93 - 0.18{\rm i}\,.
\end{eqnarray}
These dispersion relations (\ref{disperse}) are valid for ${\rm Re}(\omega)\ge 0$ for the curves I, II and III, {\it etc}, and for ${\rm Re}(\omega)\le 0$ for the primed curves.  In Fig.~\ref{disperseplot}, we show that the constructed dispersion relations fit very well with the the actual poles at least for $k$ up to $10$. The coefficients $a_{-1}, a_0$ and $a_1$ were obtained numerically. It would be of great interest to obtain their analytic expressions in terms of $r_0$ and $q$.

\noindent{\bf \large Case 2: $r_1=0$ and $r_2<r_0$.} This black hole was also obtained in \cite{luwangsusylif} and it is the non-extremal generalisation of Case 1. We label $r_0=r_+$, $r_2=r_-$ and hence $Q=r_+r_-$. We shall not present the derivation but simply the Green's function:
\begin{eqnarray}
&&G(\omega, k) = {\rm i} \fft{1 - 2{\rm i} (\gamma+ 2\Omega)}{1 + 2{\rm i} ( \gamma-2\Omega)}\,,\qquad \Omega =\fft{\omega}{4\pi T}\,,\qquad T=\ft{1}{4\pi} (r_+-r_-)\,.\cr
\gamma &=& \fft{{}_2F_1 [1-\nu - {\rm i} q,\, 1 + \nu - {\rm i} q;\,
\ft32 -2 {\rm i} \Omega;\, -\fft{r_-}{4\pi T}]\,k}{{}_2F_1[
-\nu - {\rm i} q,\, \nu - {\rm i} q;\, \ft12 -2 {\rm i} \Omega;\,
-\fft{r_-}{4\pi T}]\,(4\pi T)}\,,\quad \nu =\sqrt{\ft{k^2}Q - q^2}\,,
\end{eqnarray}
Note that $T$ is the temperature of the black hole.  Taking the subtle limit $r_-\rightarrow r_+$, we recover the results in Case 1.

For a given black hole specified by the inner and outer horizons $(r_-, r_+)$, the Fermi surfaces can be obtained by the poles of $G(0,k)$ for any given $q$. Several new properties emerge in the non-extremal case.  In the extremal limit, the general solution of $u_-$ with $\omega=0$ is given by
\begin{equation}
u_-=c_1 \Big(1-\fft{r_0}{r}\Big)^\nu + c_2 \Big(1 - \fft{r_0}{r}\Big)^{-\nu}\,.\label{um1}
\end{equation}
Thus we see that $\nu$ has to be real; otherwise, the solution becomes oscillatory on the horizon, implying instability.  In fact, there is usually a decoupling limit in an extremal black hole so that the metric becomes its near-horizon geometry AdS$_2\times S^2$. The parameter $\nu$ then measures the conformal weight of the dual operator on the AdS$_2$ boundary, and hence it must be real. Furthermore, for $\nu>0$, which we can now choose without loss of generality, we must have $c_2=0$ for the solution to be well defined on the horizon.  Thus the horizon boundary condition can be fixed for the wave solution even with $\omega=0$, for which the concept of in-falling and outgoing becomes mute.  The situation is very different for the non-extremal case.  There is no decoupling limit to obtain the horizon geometry $R^2\times T^2$. The wave solution with $\omega=0$ is not oscillatory regardless whether $\nu$ is complex or real.  In fact, as we shall see later, there is no apparent combination of parameters that can be recognised as $\nu$ in the most general back hole background.  Thus for the non-extremal case, there should be no physical requirement that $\nu$ be real, and in our case it can be pure imaginary as well.  Another new feature is that both modes of the $\omega=0$ solution on the horizon are non-divergent and hence there is no way to impose the horizon condition on the $\omega=0$ solution alone.  This adds an additional level of complexity for studying the Fermi-Dirac system at finite temperature.

\begin{figure}[ht]\center
\includegraphics[width=4.5cm]{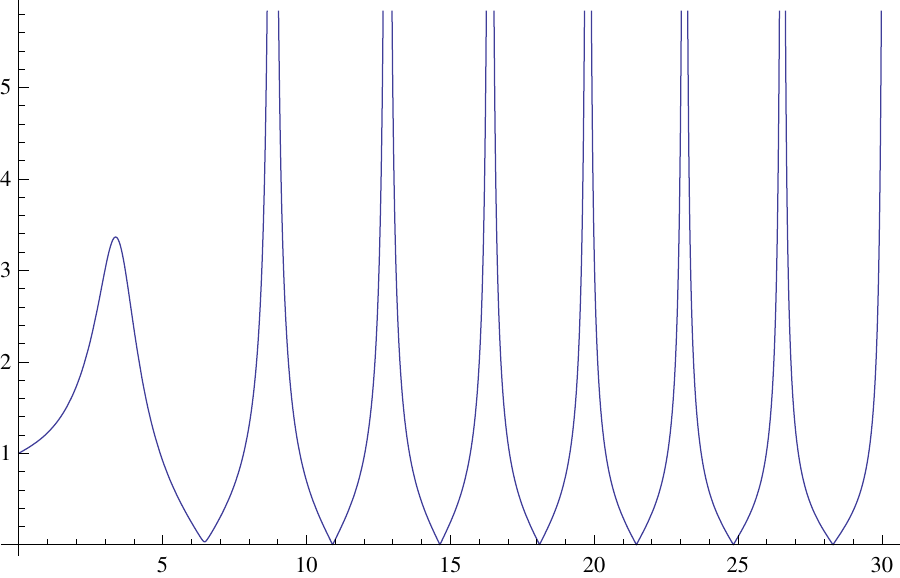}\ \ \ \
\includegraphics[width=5cm]{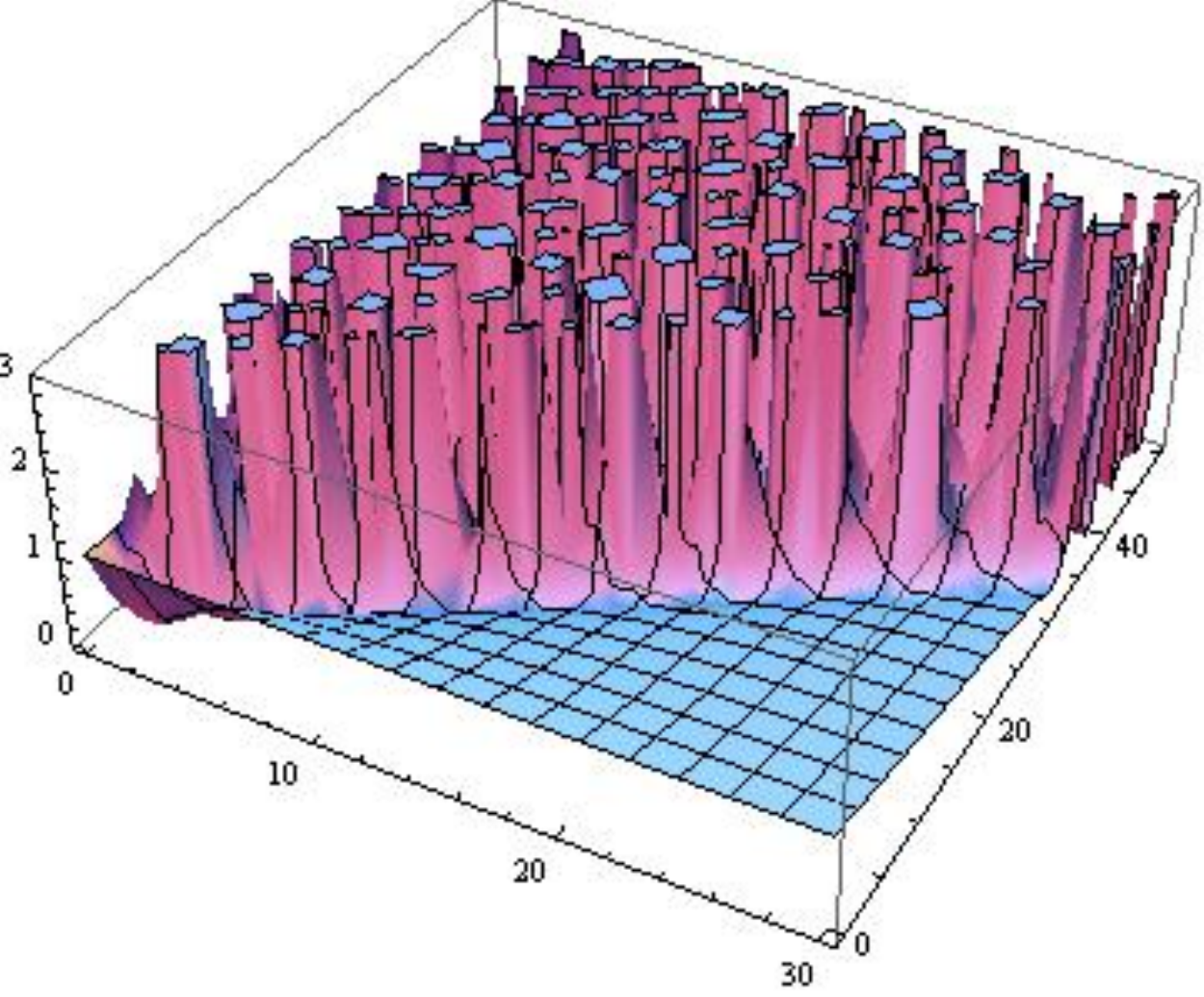}
\caption{The non-extremal black hole is specified by $r_\pm =\pm 1$.  The left plot is $|G|$ as a function of $k$ with $q=50$.  The right is a 3D plot of $|G|$ with $0\le k\le 30$ and $0\le q \le 50$.}
\label{fs}
\end{figure}

Another important new feature in the non-extremal case is that the Green's function $G(0,k)$ is not real whilst it is always real in the extremal case.  This implies that when we adjust the parameter $k$, it is unlikely that both the real and imaginary parts of $1/G(0,k)$ vanish simultaneously.  Thus the maxima of $|G(0,k)|$
are in general not literally divergent. However, we do find many examples of spiked maxima that are of orders $>10^8$. For all practical purposes, they can be viewed as divergent and the corresponding $k$'s can be regarded as Fermi surfaces. In Fig.~\ref{fs}, we present the plots of $|G(0,k)|$ and we see that spiked maxima occur; some of them are of the order $10^8$ and higher.  For example, at $k_F= 43.18346539964$ we have $|G|\sim 10^{12}$ for $r_\pm=\pm 1$ and $q=50$.

\begin{figure}[ht]\center
\includegraphics[width=7cm]{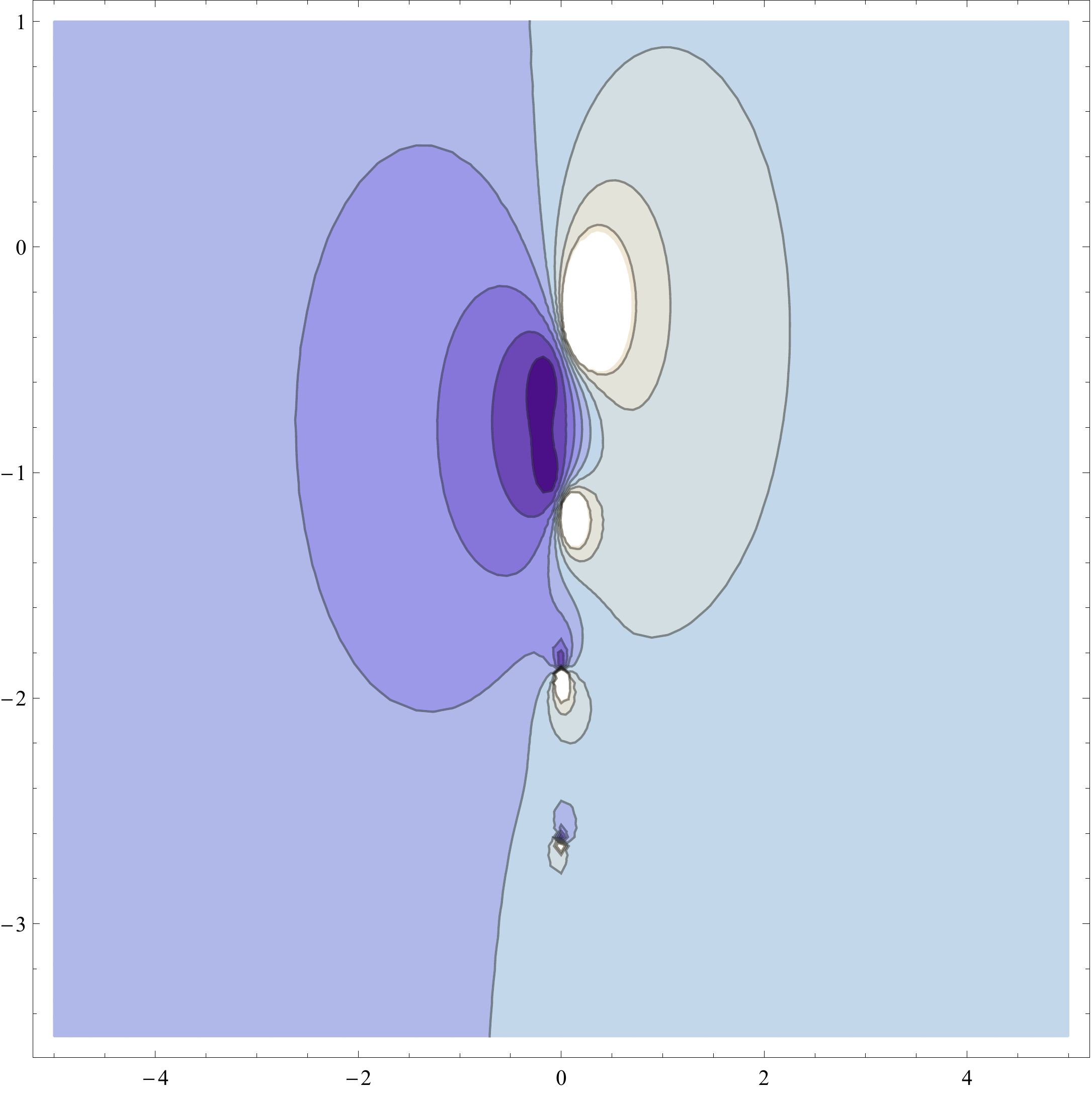}\ \ \ \
\includegraphics[width=7cm]{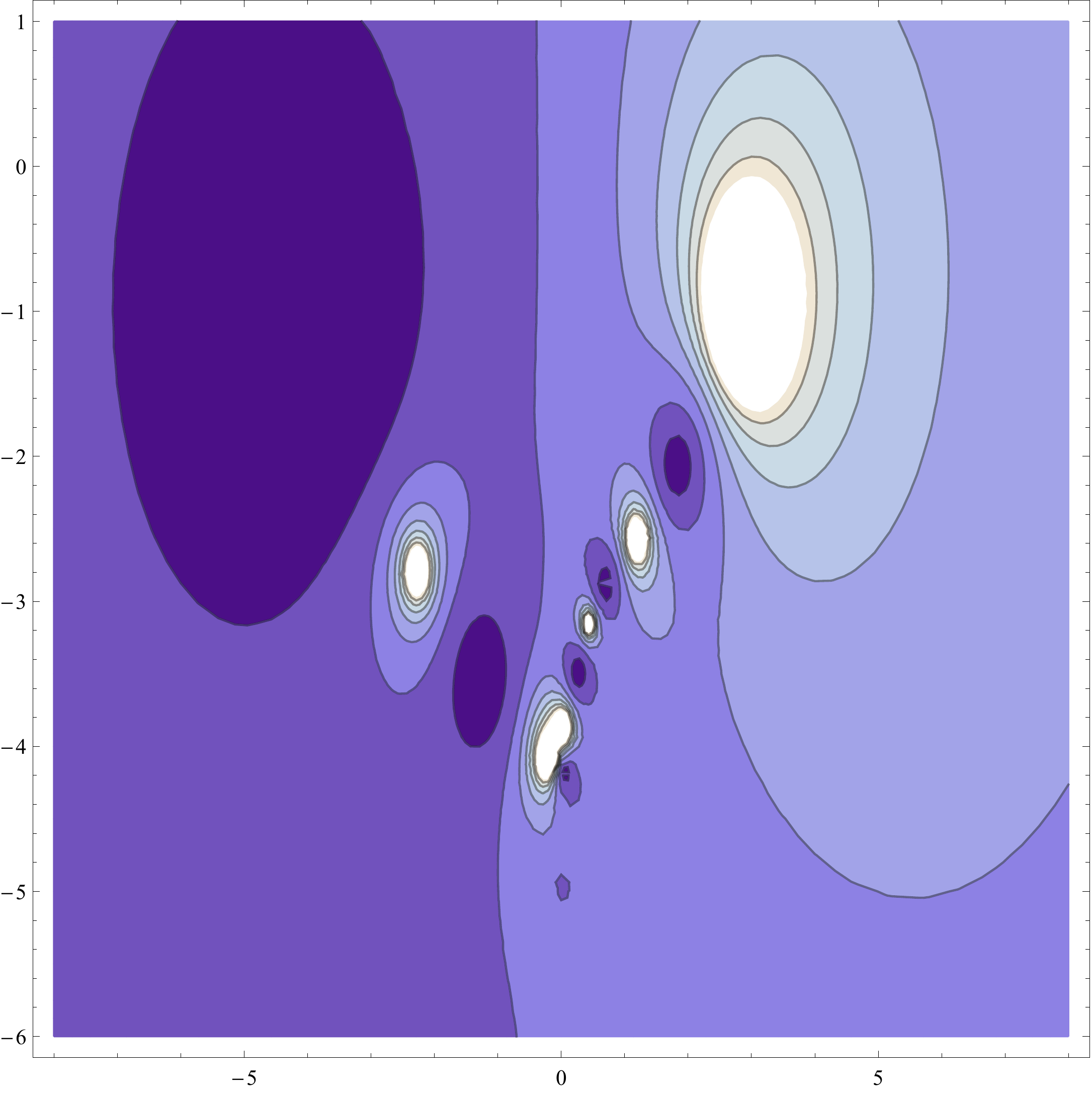}
\caption{In the non-extremal case, the patterns of zeros and poles of $|G|$ in the complex $\omega$ plane become more complicated. The black hole parameters are $r_+=2=1/r_-$ and $q=1$. The left contour plot is for $k=1/2$ with imaginary $\nu$ and the right is for $k=4$ with real $\nu$. There is no branch cut in non-extremal case.}
\label{necon}
\end{figure}

As in the case one, it is instructive to plot $|G(\omega, k)|$ in the plane of complex $\omega$, which we present in Fig.~\ref{necon}.  The black hole parameters are $r_+=2=1/r_-$ and the spinor charge is $q=1$. The pattern of zeros and poles becomes more complicated in the non-extremal case.  An important new feature is that there is no branch cut in the non-extremal case.  As we change the values of $k$, we find that new patterns of zeros and poles emerge with some isolated pairs spread in the complex plane.  The motion of the poles in the complex plane and also the dispersion relation require further investigation in a future publication.  Interestingly, we find that as the black hole approaches the extremality, some group of zeros and poles condense on the negative imaginary axis to form the branch cut.

It is of interest to understand how in the extremal limit the quantity $\nu$ is restricted to real values.  Let $r_-=r_+-\delta$. We consider $0<\delta/r_+<<1$.
We find that if we let $\nu$ to be pure imaginary, for a given $q$, no matter how small, as $\delta$ becomes smaller, there will always be a Fermi surface.
In fact there is an oscillatory factor $\delta^{\nu}$ for imaginary $\nu$ in the Green's function, implying that as $\delta$ becomes smaller and smaller, more and more Fermi surfaces emerge for $k$ lies between $0$ and $q r_0$, for which $\nu$ is pure imaginary. The maximum $k_F$ for the Fermi surfaces becomes $q r_0$ in the extremal limit.  As $\delta$ approaches zero, the number of Fermi surfaces for $k< q r_0$ becomes infinite, signalling instability.  This is the origin of the instability observed earlier from the wave function of imaginary $\nu$  that has an infinite number of oscillations near the horizon of the extremal black hole.

\noindent{\bf \large Case 3: $r_1<r_2=r_0$.} This is the general extremal solution with two non-trivial parameters.  The general $G(\omega, k)$ is of confluent Heun's functions.  The $G(0,k)$ is simpler, expressible in terms of a hypergeometric function:
\begin{eqnarray}
&&G(0,k) = \fft{{\rm Im} (Z)}{{\rm Re}(Z)}\,,\qquad \nu =\sqrt{\ft{k^2}{r_0(r_0-r_1)} - q^2}\,,\cr
Z &=& (1-{\rm i}) \Big(\nu- \fft{k}{\sqrt{r_0(r_0-r_1)}}+  {\rm i}q\Big)
\,{}_2F_1[\nu-{\rm i} q, \ft12 + \nu - {\rm i} q; 1 + 2\nu; - \fft{r_1}{r_0-r_1}]\,.
\end{eqnarray}
Note that in the extremal case, we must impose reality condition on $\nu$ and the Green's function is manifestly real. For large enough $q$, multiple fermi surfaces can emerge.  Further discussions are given in \cite{lllw}.

\noindent{\bf \large Case 4: the most general black hole.} As shown in \cite{lllw}, the general Green's function can be expressed analytically in terms of Heun's functions, namely $G(\omega, k) = {\rm i} \fft{1 - \gamma}{1 + \gamma}$, where
\begin{eqnarray}
&&\gamma= \fft{c\, H\!\ell\Big( a, \beta^*; \ft12 - 2 {\rm i} \Omega_0, 1- 2 {\rm i} \Omega_0, \ft32- 2 {\rm i} \Omega_0, \ft12 + 2{\rm i} (\widetilde Q + \Omega_1); -\fft{r_1}{r_0-r_1}\Big)}{H\!\ell\Big( a,b ; 0, \ft12, \ft12 - 2 {\rm i} \Omega_0, \ft12 - 2{\rm i} (\widetilde Q + \Omega_1); -\fft{r_1}{r_0-r_1}\Big)}\,,\cr
&&a=\fft{(r_0-r_2)r_1}{(r_0-r_1)r_2}\,,\quad  b=-\fft{k^2}{(r_0-r_1) r_2}\,,\quad
\widetilde Q= \fft{q Q}{r_0(r_1-r_2)}\,,\quad \Omega_1=\fft{\omega r_1}{(r_1-r_0) (r_1-r_2)}\,,\cr
&& \Omega_2=\fft{\omega r_2}{(r_2-r_0) (r_2-r_1)}\,,\qquad
\Omega_0=\fft{\omega r_0}{(r_0-r_1) (r_0-r_2)}=\fft{\omega}{4\pi T}\,,\cr
&&\beta=b - (\ft12 + 2{\rm i} \Omega_0)\Big(\fft{2{\rm i} q Q}{r_2(r_0-r_1)} + \fft{2r_1 r_2 - (r_1+r_2) r_0}{2 r_2 (r_0-r_1)}(1 + 4{\rm i}\Omega_0)\Big)\,,\cr
&&c=\fft{\rm i}{k} (1-\ft{r_1}{r_0})^{\fft12 -2{\rm i} (\widetilde Q + \Omega_1)}\, (1 - \ft{r_2}{r_0})^{\fft12 + 2{\rm i} (\widetilde Q - \Omega_2)}\,.\label{gengreens}
\end{eqnarray}
We shall not analyse the Green's function here; it suffices to mention that spiked maxima of $|G(0,k)|$ in the orders beyond $10^{8}$ do arise in this general case.  Details can be found in \cite{lllw}.  It is important to note that there is no parameter combination in the above that resembles $\nu$.

{\bf \large Conclusions}: In this paper we present the $T^2$-symmetric charged AdS black holes in conformal gravity.  We report that the Dirac equation of a charged massless spinor in this background can be solved exactly in terms of Heun's functions.  We give detail analysis for the simplest extremal black hole background and present some salient results of the most general situation, for which more details can be found in \cite{lllw}. Whilst the near Fermi-surface behaviour in the extremal case is expected to be similar to the previous results, our general $G(\omega, k)$ allows us to study properties beyond Fermi surfaces in analytic way.  We study the motion of the poles in the plane of complex $\omega$ and obtain the leading orders of the dispersion relation, expressed as the Laurent expansion of $\omega$ in terms of $k$, with the coefficients determined numerically.

Furthermore, we have exact $G(\omega,k)$ for non-extremal black holes as well and there is a rich structure of spiked maxima in $|G(0,k)|$.  This allows one to study some strongly interacting fermionic system at finite temperature.  We illustrated many new distinguishing features at finite temperature.  Conformal gravity with the Maxwell field exists only in four dimensions.  Our results may provide a new way of studying some two-dimensional condensed matter system such as graphene in the strongly coupled region.

\noindent{\bf \large Acknowledgement}: We are grateful to Jun Li and Hai-Shan Liu for useful discussions. H.L.~is grateful to KIAS for hospitality.  The research of H.L.~is supported in part by NSFC grants 11175269 and 11235003.

\end{document}